\documentclass[11pt,twoside]{article}
\usepackage{ldp5,flushrt,epsfig}

\begin{document}

\setcounter{page}{1}

% Put the title of your paper, in capitals, on the next line.
\title{Near--Infrared Spectroscopy and Young Stellar Populations}

\author{Michael R. Meyer }
\affil{Steward Observatory, The University of Arizona}

\GARfoot{-10}{Michael R. Meyer}
{Steward Observatory}
{The University of Arizona}{Tucson, AZ 85721}{mmeyer@as.arizona.edu}
\vspace{.3in}

\markboth{\hspace{0.5in}{\rm M.R. Meyer}
\hspace{\fill}{\rm }}{{NEAR--IR SPECTROSCOPY AND YOUNG STARS}}

\abstract{
Over the past decade, tremendous progress has been made in applying the
techniques of MK classification to spectra obtained from 1--5 $\mu$m.  
This spectral region is rich in atomic and molecular features which 
are temperature and luminosity sensitive providing a powerful 
technique to study intrinsically red and/or heavily dust embedded stellar 
populations.  We will summarize recent work with particular emphasis
on data obtained for a set of 88 fundamental MK standards in the 
J--, H--, K--, and L--bands with the KPNO 4m FTS.  One area that has
benefited enormously from these efforts has been the study of extremely
young stellar populations still embedded in the molecular cloud cores
from which they formed.  Such objects present peculiar challenges
in applying the MK process such as sub--giant surface gravities at
late--types and enhanced stellar activity associated with accretion
from circumstellar disks.  We will provide an overview of these 
difficulties and present preliminary results on possible solutions.} 
\section{{INTRODUCTION}}
\nopagebreak
Since the dawn of human conscienceness, we 
have been trying to understand our place in the Universe. 
Yet only within the past century, have astronomers realized that
by studying the formation of stars in the Milky Way galaxy, 
we can gain insight into the origin and evolution of our 
own solar system.  In order to arrive at a complete theory 
of the star formation process, we must be able to 
interpret clues provided by observations of young 
stars in the pre--main sequence (PMS) phase of evolution. 
A solar mass star requires just over 30 Myr 
to reach the main sequence once it begins its decent
down a classical Hyashi track 
taking less time for higher mass stars and 
more time for lower mass stars 
(Palla, 1999 and references
therein).  Large samples of young stars
within 1 kpc of the Sun provide the basis for a 
detailed understanding of the process of 
star--formation in our galaxy.  Also, by studying
star formation locally, we can hope to infer the
mass--to--light ratios and formation history of other 
galaxies based on observations of their integrated light. 

Over the past 30 years, astronomers interested in understanding
the star formation process have been driven to observe
at near--infrared wavelengths.  Why?  
Young stars are often found embedded
in the gas and dust from which they formed and the extinction due 
to ISM dust grains at 2.2 $\mu$m is approximately $1/10$
that observed at visible wavelengths (Rieke and Lebofsky, 1979).  Infrared
observations can enable derivation of star formation 
histories and intial mass functions 
unbiased with respect to stellar mass or youth.  Recent work 
has shown that infrared spectroscopy is a powerful tool to 
complement imaging photometric surveys of young stellar populations. 
In addition, infrared spectroscopy from 1--5 $\mu$m enables the study
of cool photopheres with T$_{eff} <$ 3000K such as M, L, and T dwarfs
(Burgasser, this volume), as well as carbon stars and Miras 
(e.g. Joyce et al. 1998).  Recent advances in detector technology 
have enabled the construction of new grating spectrographs which 
have revolutionized the study of stellar photospheres in the 
near--infrared.  We begin with a review of progress made, 
and outline the challenges in applying these techniques to 
young stellar populations.  We conclude with a brief summary 
and discussion of future directions. 
\section{{NEAR--INFRARED SPECTROSCOPY}}
\nopagebreak
It is worth noting that when Herschel (1800) 
discovered the infrared spectral region, he was
indeed observing the {\it spectrum} of the Sun!
In the first modern work on stellar near--infrared 
classification spectroscopy, 
Johnson and Mendez (1970) published a limited atlas.
These pioneering observations were made with a 
Fourier Transform Spectrometer (FTS) on Mt. Lemmon 
in southern Arizona.  Although the range of 
spectral types was not extensive, the atlas 
was a fundamental reference for many years. 
Throughout the 1970s and early 1980s, many
researchers contributed to the development of 
infrared spectroscopy (Merrill \& Ridgway, 1979). 
Given the primitive state of near--IR detectors, 
the unique properties of the FTS, which multiplexes the signal (and thus 
the noise) from all spectral elements simultaneously
on the detector, provided an advantage over 
traditional grating spectrographs (Ridgway \& Brault, 1984). 
Kleinmann and Hall (1986; hereafter KH896) provided the first 
comprehensive stellar atlas of stars in the K--band
utilizing the KPNO 4m FTS.  However, 
because it was not routine to obtain near--IR 
spectra of fainter sources to which one could
compare the standard sequence, the power of 
the atlas was not fully utilized. 

With the advent of large format, high quantum 
efficiency near--infrared arrays in the late
1980s and early 1990s, sensitive grating spectrographs
that took advantage of this revolution in detector
technology were constructed.  Prominent early work
included the following.  Kirkpatrick et al. (1993)
utilized a linear array to obtain classification 
spectra of late--type standard stars from K5--M9. 
Lancon \& Rocca--Volmerange (1992) used an FTS on the 
CFHT to produce an H-- and 
K--band atlas of stellar spectra obtained at a 
resolving power of R$=$ 400 at modest signal--to--noise. 
Oliva, Moorwood, and Origlia (1993) utilized the IRSPEC
grating spectrograph at the ESO--NTT telescope to construct 
an H--band atlas at R$=$1500 for a limited range of 
spectral types.  Ali et al. (1995) used the OSIRIS 
spectrograph on the CTIO 1.5m telescope to produce 
a K--band atlas at R$=$1400 for late--type stars. 
Hanson et al. (1996) used a variety of instruments 
to produce a K--band atlas of
early--type stars at R$\sim$1000.  With the wealth 
of new sensitive instruments on 2--4m class telescopes, 
astronomers could obtain R$=$1000 spectra of K $<$ 12.5$^m$
objects in less than one hour. 
As a result, interest in stellar infrared spectroscopy exploded! 

However, a grid of fundamental MK standards obtained 
at medium resolution with high SNR
over the full range of spectral--types and luminosity 
class, was lacking.  We undertook such a survey with 
the KPNO 4m FTS before it was decommisioned in 1997. 
Although the FTS was limited in practice to observing 
very bright stars (H $<$ 5.0$^m$ for R$>$ 3000), 
it provided certain advantages for our program
(Meyer et al. 1998; hereafter MEHS98). 
Chief among these was the ability to simultaneously subtract 
sky emission in the fourier domain enabling us
to observe with the 4m during the morning after the
night--time observers had terminated their program. 
We adopted 88 standard stars from "primary" lists 
including: Morgan, Abt, and Tabscott (1978) for O6--G0; 
Keenan \& McNeil (1989) G0--K5; and Kirkpatrick et al. 
(1991) for stars K5--M3.  We covered 26 spectral type
``bins'' and three luminosity class regimes 
(I--II, III, IV--V) of near--solar metallicity. 
Our target sample of standards
\footnote{Prof. Garrison was very gracious in engaging
in discussions during the early
phases of sample selection both during an 
April Fool's Day run at San Pedro Martir Observatory 
and electronically.} is shown in Figure 1. 
Our atlas is published in a series of papers with 
spectra covering 1--4 $\mu$m (but recorded as 
as relative flux vs. wave number) at R$=$3000 and SNR $>$ 75
\footnote{These data are available electronically at 
ftp.noao.edu/catalogs/medresIR.}. 

\placepsfig{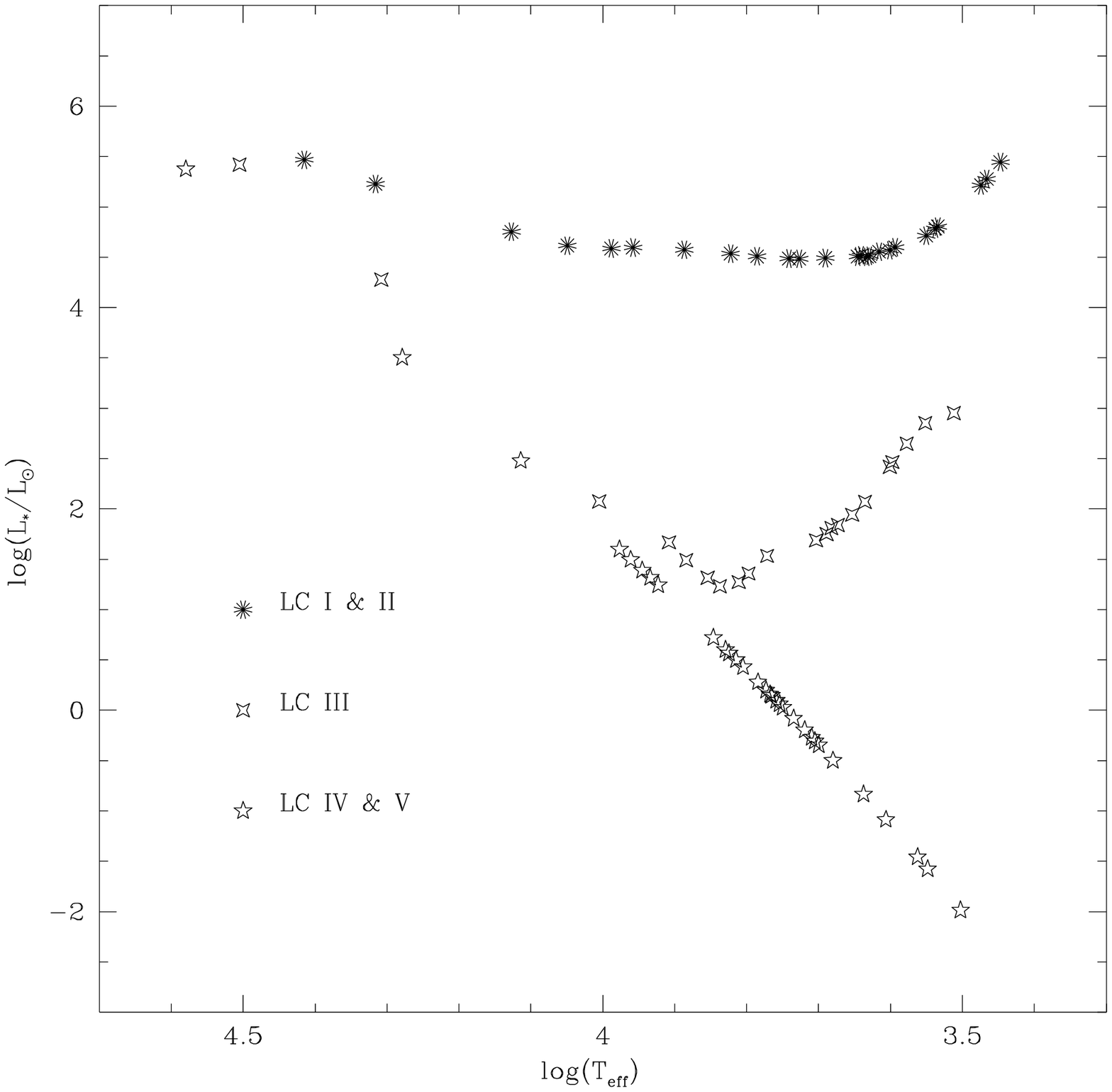}{4.00in}{0}{H--R Diagram of MK standards 
from the KPNO 4m FTS Survey (MEHS98).}

Wallace \& Hinkle (1997) published the first paper in the series, 
a high quality reference atlas that built upon the work 
of KH86 in the K--band.  The spectra were 
corrected for telluric absorption and particular attention
was paid to line identification of weak features in the spectra.  
MEHS98 presented spectra of these
standard stars in the H--band.  Equivalent widths of major 
features were measured as a function of spectral type
and luminosity class.  An attempt was made to define a 
two--dimensional classification scheme based on 
equivalent width {\it line ratios} in order to 
minimize errors in derivation of stellar properties
for stars with excess continuum emission.  Wallace, 
Meyer, Hinkle, \& Edwards (2000) published a comparable analysis 
of the J--band spectra.  Care was taken to make accurate
corrections for telluric water vapor absorption 
which was found to be time variable. 
The last of the four papers describes 
the L--band spectra (Wallace \& Hinkle, 2002; Figure 2).  We summarize
the utility of each spectral band for spectral classification
based on our experience in Section 4. 

\placepsfig{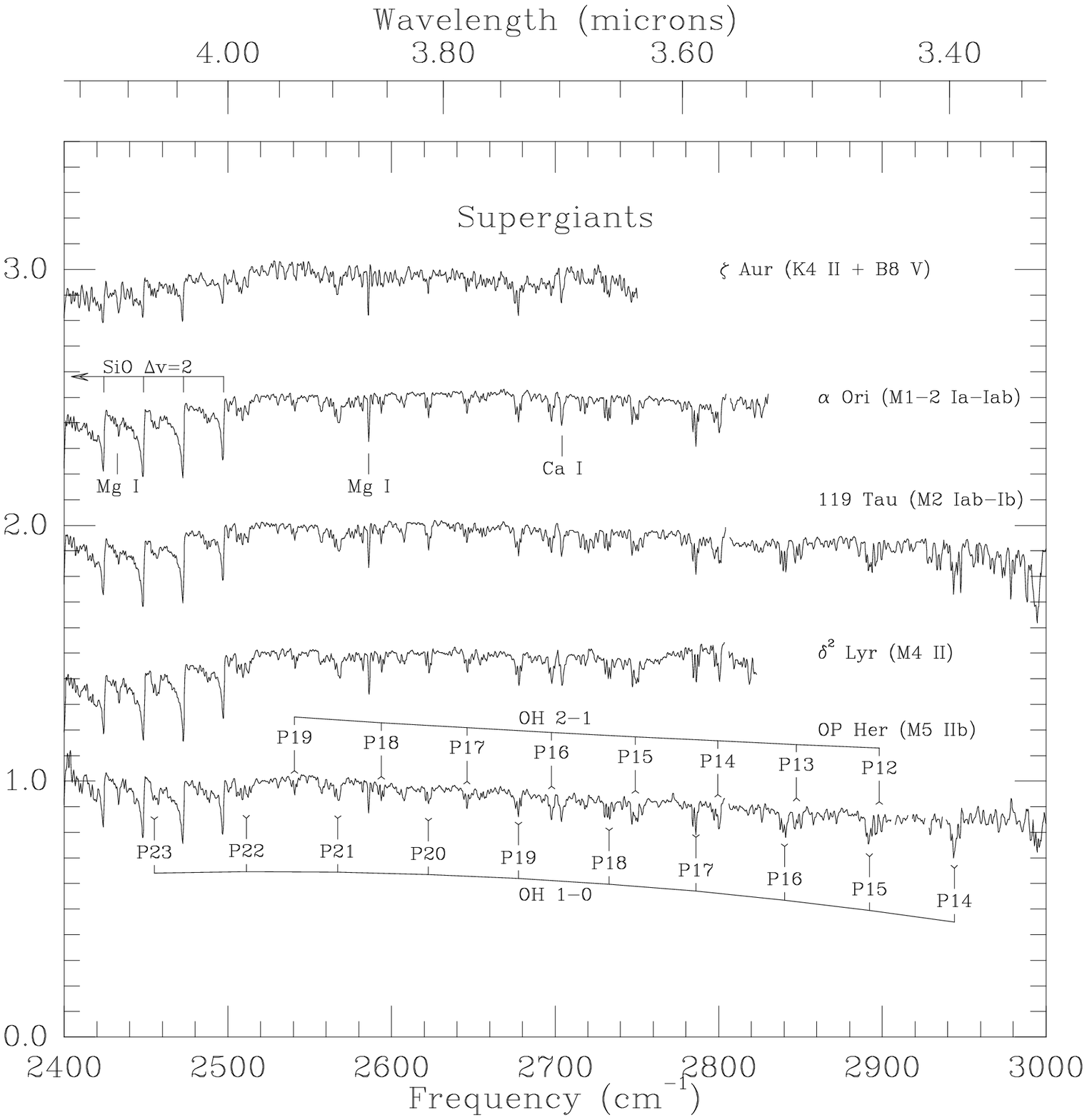}{4.00in}{0}{L--band spectra of MK standards 
obtained with the KPNO 4m FTS (Wallace \& Hinkle, 2002).}

\section{{YOUNG STELLAR POPULATIONS}}
\nopagebreak

Current work builds on a rich heritage of spectral analysis of
young stars. 
The spectroscopic study of young stellar populations began with 
the identification of the T Tauri class of objects by 
Joy (1945).   Herbig (1962) outlined issues for further study 
by reviewing the properties of the ``Orion Population'' 
of young variable stars.  Many dark clouds and star--forming 
regions are intimately associated with OB associations and HII 
regions.  Blauuw (1964) used narrow--band Walraven photometry 
to study the Gould's Belt association of young OB groups.
Garrison (1967) used traditional MK spectroscopy to study 
the young massive stars in the Sco--Cen OB association.  
Cohen and Kuhi (1979) provided the first atlas of 
digital spectra for a large number of low mass late--type
stars in a number of star--forming regions.  Through 
determination of the spectral type of a young star, 
one can infer its temperature, derive reddening 
corrections from the photometry, estimate its luminosity, 
and place the star
in a theoretical Hertzsprung--Russell Diagram.  
Comparison of the position of  young stars in this
diagram with theoretical pre--main sequence 
evolutionary tracks enables one to estimate the 
mass and age of the objects in question.  Such estimates 
provide the fundamental data needed 
to study the initial mass function
of a young stellar population and 
the star forming history of a molecular cloud
(e.g. Hillenbrand, 1997). 

What are the special problems one faces
when trying to study young stars spectroscopically? 
There are three main problems that can hamper 
efficient determination of spectral types from 
``classification resolution'' (500 $< R <$ 1000) 
spectra.  First, very young stars are often associated
with the clouds of gas and dust from which they are 
created.  As a result, extinction can range from 
a few to tens of magnitudes in the V--band 
\footnote{Prof. Garrison (1968) helped initiate attempts 
to derive MK types for heavily reddened stars, such as
IRS \#1 in NGC 2024, the flame nebula.}.  This 
complicates derivation of spectral types that rely
on colors or broad--band indices, unless a reddening--free 
index can be used.  Second, 
by their very nature young low mass stars have not yet 
descended completely down their convective 
tracks.  As a result, their radii are large and 
their surface gravities 
are lower ($3.0 < log(g) < 4.0$) than their 
main sequence counter--parts ($4.0 < log(g) < 5.0$)
as illustrated in Figure 3. 
This complicates derivation of spectral types through 
comparison of program stars with main sequence 
standard stars because for stars earlier than M0, 
a given spectral type (i.e. ionization state) 
corresponds to a cooler photosphere at lower 
gravity (Gray, 1991).  However, for stars later than M0, lower
surface gravity photospheres are {\it hotter} than 
high surface gravity counter--parts for a given spectral type
(c.f. Perrin et al. 1998; Wilking et al. 1999; 
see also Itoh et al. 2002, their Figure 4). 
Unfortunately, our Milky Way galaxy has not yet had time to 
produce evolved sub--giant ``standard stars'' later than K2. 
Further more, intrinsic colors and bolometric 
corrections can also be a function of surface gravity, 
introducing additional uncertainties into luminosity estimates 
for young stars (e.g. Bessell et al. 1998).  
Finally, excess emission associated with ``activity'' can
veil spectral features in at least two wavelength regimes.
The impact points of accretion streams onto the central 
stars can emit as pseudo--blackbodies at temperatures 
well in excess of the photosphere introducing blue 
continuum excess at wavelengths $<$ 0.5 $\mu$m (e.g. Gullbring et al. 2000).  
Warm dust ($T_{dust} < T_{sublimation}$) in circumstellar disks 
can also dilute photospheric emission in the infrared 
(e.g. Meyer et al. 1997; Johns--Krull \& Valenti, 2001). 
Typical interstellar dust sublimates at temperatures $>$ 1200--2000 K
restricting the problem to wavelengths $>$ 1.5 $\mu$m in most cases. 

\placepsfig{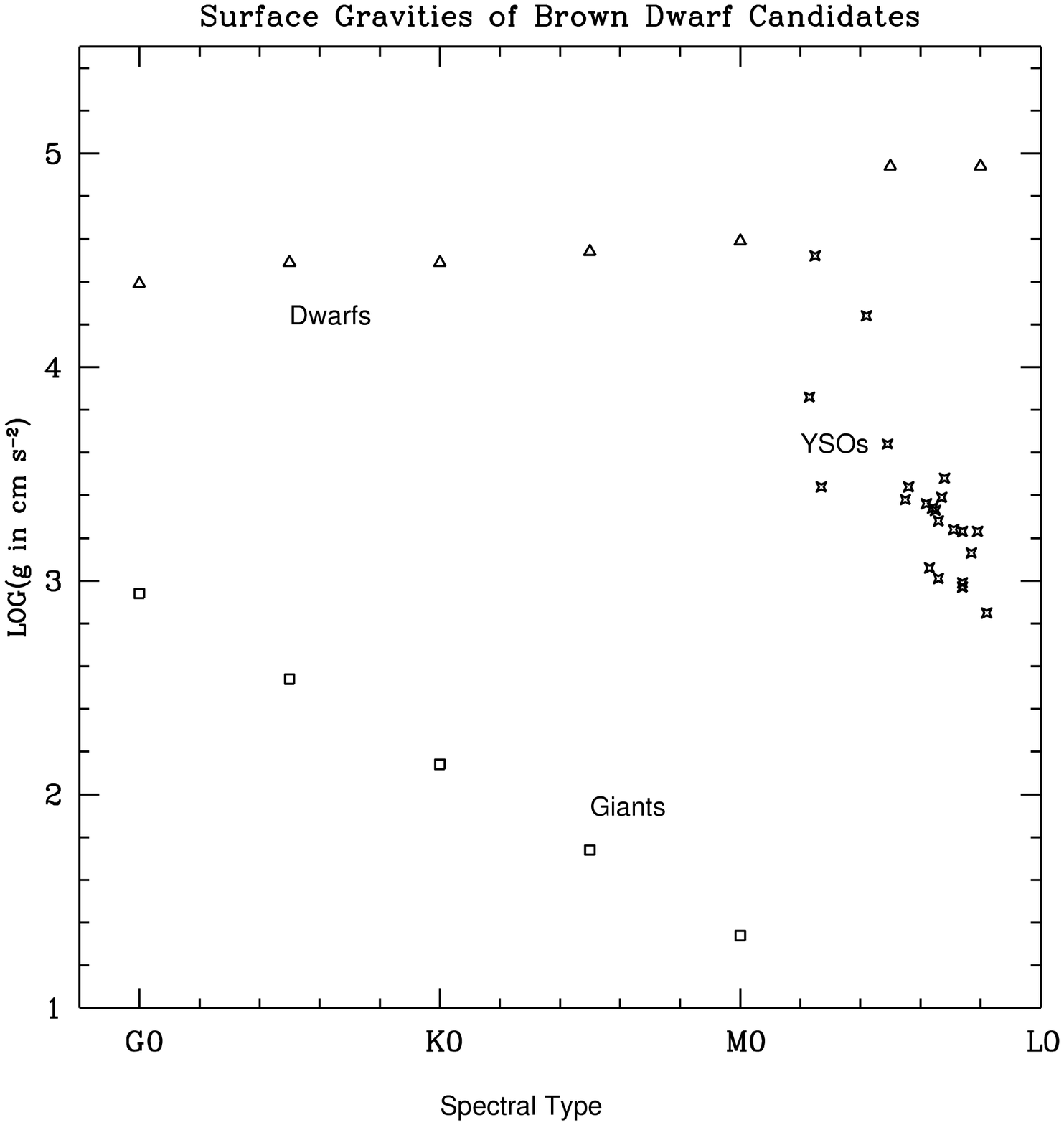}{4.00in}{0}{Surface Gravities of young late--type 
stars derived from placing the stars in the H--R diagram and comparing with 
theoretical tracks (Wilking et al. 2003).}

What are some possible solutions to these problems?  With regard
to extinction, near--infrared spectroscopy has demonstrated to be
a powerful tool for deriving MK spectral types as described above. 
Many researchers now routinely use infrared spectroscopy to study 
young stellar populations in star--forming regions 
(e.g. Carpenter et al. 1997; Luhman et al. 1998; Luhman \& Rieke, 1999; 
Wilking et al. 
1999; Meyer \& Wilking 2003).  With regard to surface gravity, 
more work is needed to understand the magnitude of the problem.
Certain diagnostics such the Sr II 0.4077 $\mu$m to Fe I 
0.4071 $\mu$m ratio (e.g. Mamajek et al. 2002), 
the NaI doublets at 0.589, 0.819 and 2.206  $\mu$m 
(Martin et al. 1996; Figure 4), 
the KI doublet at 1.22 $\mu$m (Figure 4), as well as molecular bands
such as CaH at 0.7 $\mu$m (Allen \& Strom, 1995; Wilking et al. 2003)
and the prominent first and second overtones bands of CO
in the K-- and H--bands respectively (KH86; MEHS98)
are surface gravity sensitive.  As theoretical calculations improve, 
comparison of observed spectra with model atmospheres for late--type
pre--main sequence stars may allow us to derive physical 
parameters for objects of interest (e.g. Allard et al. 2000).  
For now, we can 
develop emperical comparisons between giant, sub--giant, and 
dwarf stars, using surface gravity sensitive features as 
a criterion for cluster membership (e.g. Gorlova et al. 2003).  
Larger samples of 
eclipsing, spectroscopic, and astrometric binaries are needed in order 
to constrain radii, temperatures, and masses for pre--main 
sequence stars in various parts of the H--R diagram for 
comparison with model tracks (e.g. Steffen et al. 2001; 
Tamazain et al. 2002).  Regarding the problem of 
excess emission in the blue and infrared wavelengths, 
line ratio techniques can be used to mitigate these 
problems (MEHS98).  Furthermore, detailed 
spectrophotometry has been used successfully to ``deveil'' photospheric
spectra (e.g. Hartigan et al. 1995).  This process requires excellent 
observing conditions and is very sensitive to the 
grid of template ``standards'' obtained as a function
of temperature, surface gravity, and possibly metallicity. 

\placepsfig{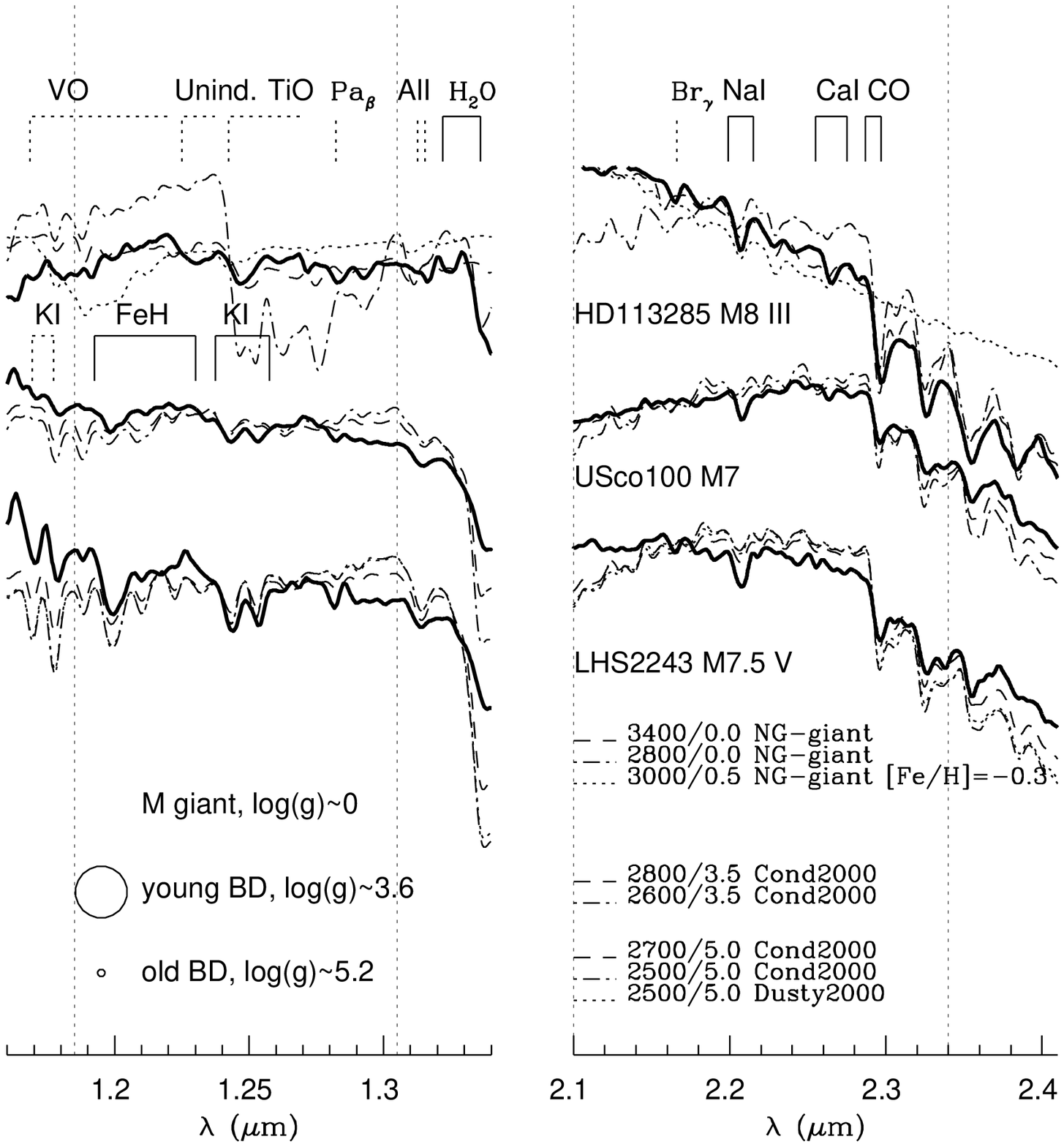}{4.00in}{0}{J--band spectra of late--type 
stars illustrating the surface gravity sensitivity of 
the KI doublet at 1.22 $\mu$m (Gorlova et al. 2003).}

\section{{SUMMARY AND FUTURE DIRECTIONS}}
\nopagebreak

In summary, near--infrared stellar spectroscopy has definitely 
``come of age'' in the last decade.  Our experiences with the 
KPNO 4m FTS MK Spectral Standards survey suggest the following:
1) The J--band, exhibiting a wealth 
of features from low ionization potential/dissociation energy atomic
and molecular species, is excellent for classifying the latest type 
stars ($>$ M3).  
Additional work is required to determine 
whether the Y--band (Hillenbrand et al. 2002) or even the 
traditional I--band is the optimum 
spectral regime for the classification of the very coolest
stars and brown dwarfs (Burgasser, this volume). 
2)  The H--band (1.5--1.8 $\mu$m) is best for intermediate
spectral types F0--K3.  There are a large number of neutral 
metal lines for classification. 
3) The K--band (2.05--2.35 $\mu$m) is very good for classifying
late--type stars K3--M3.  However, warm dust can contaminate
the spectra of PMS and evolved stars so caution is required. 
4) The L--band (3.4--3.8 $\mu$m) is not the ideal wavelength regime for 
classifying stars.  There are fewer stellar features
available over a wide range of temperatures and surface 
gravities and it is very difficult to obtain the required
signal--to--noise from ground--based observations. 

In addition to the advances in infrared spectroscopy, 
there are many exciting directions for future studies 
of young stellar populations.  Work must 
continue to develop emperical estimates of log(g) which can 
be used as membership criteria to ellimate contamination of
foreground and background stars from statistical samples. 
We need to understand the effects of surface gravity on 
sub--giant stellar parameters through comparison of models
with observational data.  Ultimately, large samples of 
PMS binaries will be required to constrain theoretical 
models.  Finally, optical and infrared multi--object spectroscopy 
can be used to create much larger spectroscopic samples
of young stars.  Perhaps the development of narrow--band
indices for very late--type stars could extend these
methods to large area filter--photometric surveys. 

It is clear that our work is not yet done. 
Viva el MK Process y MK System! Orale! 

\acknowledgments
The author would like to express his grattitude to the 
meeting organizers and proceeding editors for the
opportunity to address the conference and contribute this
manuscript, as well as to Prof. Garrison for generously 
sharing his time with junior scientists whenever the 
opportunity arises.  The author would also like to thank 
Nadja Gorlova, Ken Hinkle, 
Wilson Liu, Eric Mamajek, Lloyd Wallace, and Bruce Wilking 
for providing material used in this presentation as well
as their continued collaboration.  Special thanks to 
L. A. Hillenbrand for comments on a draft
of this manuscript.  The author is very grateful for support 
through the Lucas Foundation, and NASA grants 
HF--01098.01--97A and GF--7417 awarded by the Space Telescope 
Science Institute which is operated by the Association of Universities 
for Research in Astronomy, Inc., for NASA under contract NAS 5--26555
during the period when most of this work was carried out.

\section{{Question and Answer}}

T. von Hippel:  Can you comment on IR classification outside the
ground-based windows, e.g. with SIRTF?

M. Meyer:  In the near-IR, it would be extremely valuable to
observe in between the ground-based windows to characterize
stellar photospheric absorptions due to water vapor etc. continuously
from 1-4 microns.  However, at thermal infrared wavelengths, there
are fewer features of interest (CO fundamental, SiO) 
for stellar classification.  There have been some attempts
to extend classification schemes to longer wavelengths using
ISO (Kraemer et al. 2002; Lenorzer et al. 2002) and the SIRTF
IRS will produce R=100 spectra from 5-40 microns and R=600
spectra from 10-37 microns.

R. Garrison:  It's true that VO and TiO are very sensitive to Teff,
but the IR is missing certain atomic and electronic transitions,
whereas the blue has everything.

M. Meyer:  The IR is certainly the only option for large classes
of interesting, heavily reddened objects.  However, I agree that
whenever possible I would rather classify stars at wavelengths
just blueward of the Wien-peak for each photosphere where features
are strongest. 

Percy:  Are you concerned that the spectra of this young
population might be affected by the high "activity", rotation,
and/or magnetic fields which are associated with young stars,
but not with post-MS stars of similar temperature and gravity?

M. Meyer:  Certainly we should be concerned with all of
these effects.  At present our focus is on the effects of
sub-giant surface gravity where we do not have post--MS 
``standards''.  We think we can overcome the
difficulties associated with accretion activity and rotation.  
Recent work by Johns-Krull and Valenti (1999) 
is aimed at deriving magnetic field 
strengths of young stars. 

J. Roundtree:  What exactly are the limits of the
short-wave IR band you are calling z or Y?
Frank Low defined a Z-band years ago as a window
around 40 microns.  He called it Z because he thought
it was the longest IR wavelength that could be observed
through the Earth's atmosphere.

M. Meyer:  The half-peak transmission limits are approximately
0.96-1.12 microns.  See the paper by Hillenbrand et al. (2002).

B. Weaver:  I am surprised that your luminosity error bars are
much smaller than your temperature errors bars on your
H-R diagrams.  Why don't the distance, reddening, and
B.C. correction errors dominate the errors compared to well-
calibrated temperatures?

M. Meyer:  Formally propagating the uncertainties in luminosity
usually lead to errors of +- 0.2 dex and in temperature,
the uncertainties in temperature scales limit us to
errors of +- 300 K.


\begin{references}

\bibitem[Ali et al.(1995)]{1995AJ....110.2415A} Ali, B., Carr, J.~S., 
Depoy, D.~L., Frogel, J.~A., \& Sellgren, K.\ 1995, \aj, 110, 2415 

\bibitem[Allard, Hauschildt, \& Schweitzer(2000)]{2000ApJ...539..366A} 
Allard, F., Hauschildt, P.~H., \& Schweitzer, A.\ 2000, \apj, 539, 366 

\bibitem[Allen \& Strom(1995)]{1995AJ....109.1379A} Allen, L.~E.~\& Strom, 
K.~M.\ 1995, \aj, 109, 1379 

\bibitem[Bessell, Castelli, \& Plez(1998)]{1998A&A...333..231B} Bessell, 
M.~S., Castelli, F., \& Plez, B.\ 1998, \aap, 333, 231 

\bibitem[Blaauw(1964)]{1964ARA&A...2..213B} Blaauw, A.\ 1964, \araa, 2, 213 

\bibitem[Carpenter et al.(1997)]{1997AJ....114..198C} Carpenter, J.~M., 
Meyer, M.~R., Dougados, C., Strom, S.~E., \& Hillenbrand, L.~A.\ 1997, \aj, 
114, 198 

\bibitem[Cohen \& Kuhi(1979)]{1979ApJS...41..743C} Cohen, M.~\& Kuhi, 
L.~V.\ 1979, \apjs, 41, 743 

\bibitem[Garrison(1967)]{1967ApJ...147.1003G} Garrison, R.~F.\ 1967, \apj, 
147, 1003 

\bibitem[Garrison(1968)]{1968PASP...80...20G} Garrison, R.~F.\ 1968, \pasp, 80, 20 

\bibitem[Gorlova(2003)]{2003Gorlova} Gorlova, N. et al. 2003, in preparation (see astro-ph/0208221). 

\bibitem[Gray 1991]{1991Gray} Gray, D.F. Observation and 
Analysis of Stellar Photospheres, (New York: Cambridge)

\bibitem[Gullbring, Calvet, Muzerolle, \& 
Hartmann(2000)]{2000ApJ...544..927G} Gullbring, E., Calvet, N., Muzerolle, 
J., \& Hartmann, L.\ 2000, \apj, 544, 927 

\bibitem[Hanson, Conti, \& Rieke(1996)]{1996ApJS..107..281H} Hanson, M.~M., 
Conti, P.~S., \& Rieke, M.~J.\ 1996, \apjs, 107, 281 

\bibitem[Hartigan, Edwards, \& Ghandour(1995)]{1995ApJ...452..736H} 
Hartigan, P., Edwards, S., \& Ghandour, L.\ 1995, \apj, 452, 736 

\bibitem[Herbig(1962)]{1962AdA&A...1...47H} Herbig, G.~H.\ 1962, Advances 
in Astronomy and Astrophysics, 1, 47 

\bibitem[Herschel(1800)]{1800RSPS....1...20H} Herschel, W.\ 1800, Royal 
Society of London Proceedings Series I, 1, 20 

\bibitem[Hillenbrand(1997)]{1997AJ....113.1733H} Hillenbrand, L.~A.\ 1997, 
\aj, 113, 1733 

\bibitem[Hillenbrand, Foster, Persson, \& 
Matthews(2002)]{2002PASP..114..708H} Hillenbrand, L.~A., Foster, J.~B., 
Persson, S.~E., \& Matthews, K.\ 2002, \pasp, 114, 708 

\bibitem[Itoh, Tamura, \& Tokunaga(2002)]{2002PASJ...54..561I} Itoh, Y., 
Tamura, M., \& Tokunaga, A.~T.\ 2002, \pasj, 54, 561 

\bibitem[Johns-Krull, Valenti, \& Koresko(1999)]{1999ApJ...516..900J} 
Johns-Krull, C.~M., Valenti, J.~A., \& Koresko, C.\ 1999, \apj, 516, 900 

\bibitem[Johns-Krull \& Valenti(2001)]{2001ApJ...561.1060J} Johns-Krull, 
C.~M.~\& Valenti, J.~A.\ 2001, \apj, 561, 1060 

\bibitem[Johnson \& Mendez(1970)]{1970AJ.....75..785J} Johnson, H.~J.~\& 
Mendez, M.~E.\ 1970, \aj, 75, 785 

\bibitem[Joy(1945)]{1945ApJ...102..168J} Joy, A.~H.\ 1945, \apj, 102, 168 

\bibitem[Joyce et al.(1998)]{1998AJ....116.2520J} Joyce, R.~R., Hinkle, 
K.~H., Wallace, L., Dulick, M., \& Lambert, D.~L.\ 1998, \aj, 116, 2520 


\bibitem[Keenan \& McNeil(1989)]{1989ApJS...71..245K} Keenan, P.~C.~\& 
McNeil, R.~C.\ 1989, \apjs, 71, 245 

\bibitem[Kirkpatrick, Henry, \& McCarthy(1991)]{1991ApJS...77..417K} 
Kirkpatrick, J.~D., Henry, T.~J., \& McCarthy, D.~W.\ 1991, \apjs, 77, 417 

\bibitem[Kirkpatrick et al.(1993)]{1993ApJ...402..643K} Kirkpatrick, J.~D., 
et al. 1993, \apj, 402, 643 

\bibitem[Kleinmann \& Hall(1986)]{1986ApJS...62..501K} Kleinmann, S.~G.~\& 
Hall, D.~N.~B.\ 1986, \apjs, 62, 501 (KH86) 

\bibitem[Kraemer, Sloan, Price, \& Walker(2002)]{2002ApJS..140..389K} 
Kraemer, K.~E., Sloan, G.~C., Price, S.~D., \& Walker, H.~J.\ 2002, \apjs, 
140, 389 

\bibitem[Lancon \& Rocca-Volmerange(1992)]{1992A&AS...96..593L} Lancon, 
A.~\& Rocca-Volmerange, B.\ 1992, \aaps, 96, 593 

\bibitem[Lenorzer et al.(2002)]{2002A&A...384..473L} Lenorzer, A., 
Vandenbussche, B., Morris, P., de Koter, A., Geballe, T.~R., Waters, 
L.~B.~F.~M., Hony, S., \& Kaper, L.\ 2002, \aap, 384, 473 

\bibitem[Luhman, Rieke, Lada, \& Lada(1998)]{1998ApJ...508..347L} Luhman, 
K.~L., Rieke, G.~H., Lada, C.~J., \& Lada, E.~A.\ 1998, \apj, 508, 347 

\bibitem[Luhman \& Rieke(1999)]{1999ApJ...525..440L} Luhman, K.~L.~\& 
Rieke, G.~H.\ 1999, \apj, 525, 440 

\bibitem[Mamajek, Meyer, \& Liebert(2002)]{2002AJ....124.1670M} Mamajek, 
E.~E., Meyer, M.~R., \& Liebert, J.\ 2002, \aj, 124, 1670 

\bibitem[Martin, Rebolo, \& Zapatero-Osorio(1996)]{1996ApJ...469..706M} 
Martin, E.~L., Rebolo, R., \& Zapatero-Osorio, M.~R.\ 1996, \apj, 469, 706 

\bibitem[Merrill \& Ridgway(1979)]{1979ARA&A..17....9M} Merrill, K.~M.~\& 
Ridgway, S.~T.\ 1979, \araa, 17, 9 

\bibitem[Meyer, Calvet, \& Hillenbrand(1997)]{1997AJ....114..288M} Meyer, 
M.~R., Calvet, N., \& Hillenbrand, L.~A.\ 1997, \aj, 114, 288 

\bibitem[Meyer, Edwards, Hinkle, \& Strom(1998)]{1998ApJ...508..397M} 
Meyer, M.~R., Edwards, S., Hinkle, K.~H., \& Strom, S.~E.\ 1998, \apj, 508, 
397 (MEHS98) 

\bibitem[Meyer(2003)]{2003Meyer} Meyer, M.R. \& Wilking, B.A. 2003, in preparation. 

\bibitem[Morgan, Abt, \& Tapscott(1978)]{1978rmsa.book.....M} Morgan, 
W.~W., Abt, H.~A., \& Tapscott, J.~W.\ 1978, Williams Bay: Yerkes 
Observatory, and Tucson: Kitt Peak National Observatory, 1978

\bibitem[Origlia, Moorwood, \& Oliva(1993)]{1993A&A...280..536O} Origlia, 
L., Moorwood, A.~F.~M., \& Oliva, E.\ 1993, \aap, 280, 536 

\bibitem[Palla(1999)]{1999osps.conf..375P} Palla, F.\ 1999, NATO ASIC 
Proc.~540: Origin of Stars and Planetary Systems, 375 

\bibitem[Perrin et al.(1998)]{1998A&A...331..619P} Perrin, G., Coude Du 
Foresto, V., Ridgway, S.~T., Mariotti, J.-M., Traub, W.~A., Carleton, 
N.~P., \& Lacasse, M.~G.\ 1998, \aap, 331, 619 

\bibitem[Ridgway \& Brault(1984)]{1984ARA&A..22..291R} Ridgway, S.~T.~\& 
Brault, J.~W.\ 1984, \araa, 22, 291 

\bibitem[Rieke \& Lebofsky(1985)]{1985ApJ...288..618R} Rieke, G.~H.~\& 
Lebofsky, M.~J.\ 1985, \apj, 288, 618 

\bibitem[Steffen et al.(2001)]{2001AJ....122..997S} Steffen, A.~T.~et al.\ 
2001, \aj, 122, 997 

\bibitem[Tamazian, Docobo, White, \& Woitas(2002)]{2002ApJ...578..925T} 
Tamazian, V.~S., Docobo, J.~;., White, R.~J., \& Woitas, J.\ 2002, \apj, 
578, 925 

\bibitem[Wallace \& Hinkle(1997)]{1997ApJS..111..445W} Wallace, L.~\& 
Hinkle, K.\ 1997, \apjs, 111, 445 

\bibitem[Wallace, Meyer, Hinkle, \& Edwards(2000)]{2000ApJ...535..325W} 
Wallace, L., Meyer, M.~R., Hinkle, K., \& Edwards, S.\ 2000, \apj, 535, 325 

\bibitem[Wallace \& Hinkle(2002)]{2002AJ....124.3393W} Wallace, L.~\& 
Hinkle, K.\ 2002, \aj, 124, 3393 

\bibitem[Wilking, Greene, \& Meyer(1999)]{1999AJ....117..469W} Wilking, 
B.~A., Greene, T.~P., \& Meyer, M.~R.\ 1999, \aj, 117, 469 

\bibitem[Wilking(2003)]{2003Wilking}Wilking, B.~A. et al. 2003, in preparation (see astro-ph/0208264) 

\end{references}
\end{document}